\documentclass[10pt,journal]{IEEEtran}
\usepackage{epsfig,rotating,setspace,latexsym,amsmath,epsf,amssymb,amsfonts,bm,theorem,cite,authblk, bbm,color, hyperref, multirow, caption, subcaption}
\IEEEoverridecommandlockouts
\allowdisplaybreaks

\begin{document}
\title{Task-Oriented Communications for NextG: \\ End-to-End Deep Learning and AI Security Aspects
}

\author[1]{Yalin E. Sagduyu}
\author[2]{Sennur Ulukus}
\author[3]{Aylin Yener}

\affil[1]{\normalsize  Virginia Tech, Arlington, VA, USA}

\affil[2]{\normalsize University of Maryland, College Park, MD, USA}

\affil[3]{\normalsize  The Ohio State University, Columbus, OH, USA}	

\maketitle

\begin{abstract}
Communications systems to date are primarily designed with the goal of reliable transfer of digital sequences (bits). Next generation (NextG) communication systems are beginning to explore shifting this design paradigm to reliably executing a given task such as in task-oriented communications. In this paper, wireless signal classification is considered as the task for the NextG Radio Access Network (RAN), where edge devices collect wireless signals for spectrum awareness and communicate with the NextG base station (gNodeB) that needs to identify the signal label. Edge devices may not have sufficient processing power and may not be trusted to perform the signal classification task, whereas the transfer of signals to the gNodeB may not be feasible due to stringent delay, rate, and energy restrictions. Task-oriented communications is considered by jointly training the transmitter, receiver and classifier functionalities as an encoder-decoder pair for the edge device and the gNodeB. This approach improves the accuracy compared to the separated case of signal transfer followed by classification. Adversarial machine learning poses a major security threat to the use of deep learning for task-oriented communications. A major performance loss is shown when backdoor (Trojan) and adversarial (evasion) attacks target the training and test processes of task-oriented communications.
\end{abstract}


\section{Introduction}
The goal of \emph{conventional communications} is to transfer information from sources to destinations. To achieve this goal, communication functions are traditionally divided into separate blocks that are individually designed, namely \emph{source coding}, \emph{channel coding}, and \emph{modulation} on the transmitter side and \emph{source decoding}, \emph{channel decoding}, and \emph{demodulation} on the receiver side.  The performance is measured by the bit/symbol error rate or a signal distortion metric, such as mean squared error, to quantify the fidelity of information reconstruction at the destination. It is possible to improve this performance with the joint design of communication functions, e.g., \emph{joint source-channel coding} \cite{bourtsoulatze2019deep}, that can be trained as deep neural networks (DNNs) to learn from and adapt to the channel effects. In addition, transmitter functions of channel coding and modulation and the corresponding receiver functions can be jointly optimized as an \emph{autoencoder} for end-to-end communications \cite{Oshea1}.

Conventional communications ignores the semantics of information by assuming that all symbols/bits are created and treated equal, while the goal remains to be the high fidelity recovery of symbols/bits at destinations. Next generation (NextG) communication systems are finding applications to challenge this traditional design paradigm. For example, \emph{semantic communications} has emerged as a new paradigm to reliably communicate the meanings of messages (instead of symbols) by minimizing the semantic error \cite{guler2014semantic} to best preserve the meaning of recovered messages. 
Another game changing approach is to consider the significance of information transfer that is pertinent to an underlying task \cite{uysal2021semantic}. Accordingly, \emph{task-oriented communications} aims to optimize the communication functions in order to perform a task (e.g., classification of signals captured by an edge device) that is the main reason why communication functions (e.g., wireless connections from the edge devices to a decision center) are needed in the first place. 

Machine learning has been considered a key enabler for \emph{NextG communications systems}. Starting with 5G, key complex tasks are envisioned to be solved by machine learning, including resource allocation and admission control for network slicing, massive MIMO optimization, channel estimation/tracking, dynamic spectrum sharing, user equipment (UE) identification and authentication. In previous generations of communications systems, rule based solutions have been primarily adopted that cannot meet the growing demand for communications and computation resources. In addition, NextG communications aims to connect edge devices/sensors and provide new services with additional computational needs such as in autonomous driving that benefit from the use of machine learning. To that end, NextG communication systems are envisioned to be \emph{task-oriented}. Starting with 5G, meeting the quality of experience (QoE) needs has become the main objective in NextG communications systems that are designed to serve tasks for diverse applications (ranging from virtual/augmented reality to autonomous driving). These tasks are orchestrated as network slices in the radio access network (RAN), such as enhanced Mobile Broadband (eMBB), massive machine-type communications (mMTC), and ultra-reliable low-latency communications (urLLC)). Therefore, task-oriented communications has emerged as a natural approach to perform the emerging tasks involved in NextG communications systems  \cite{strinati20216g, xu2022edge}. 

\begin{figure}[t]
\centering
\includegraphics[width=\columnwidth]{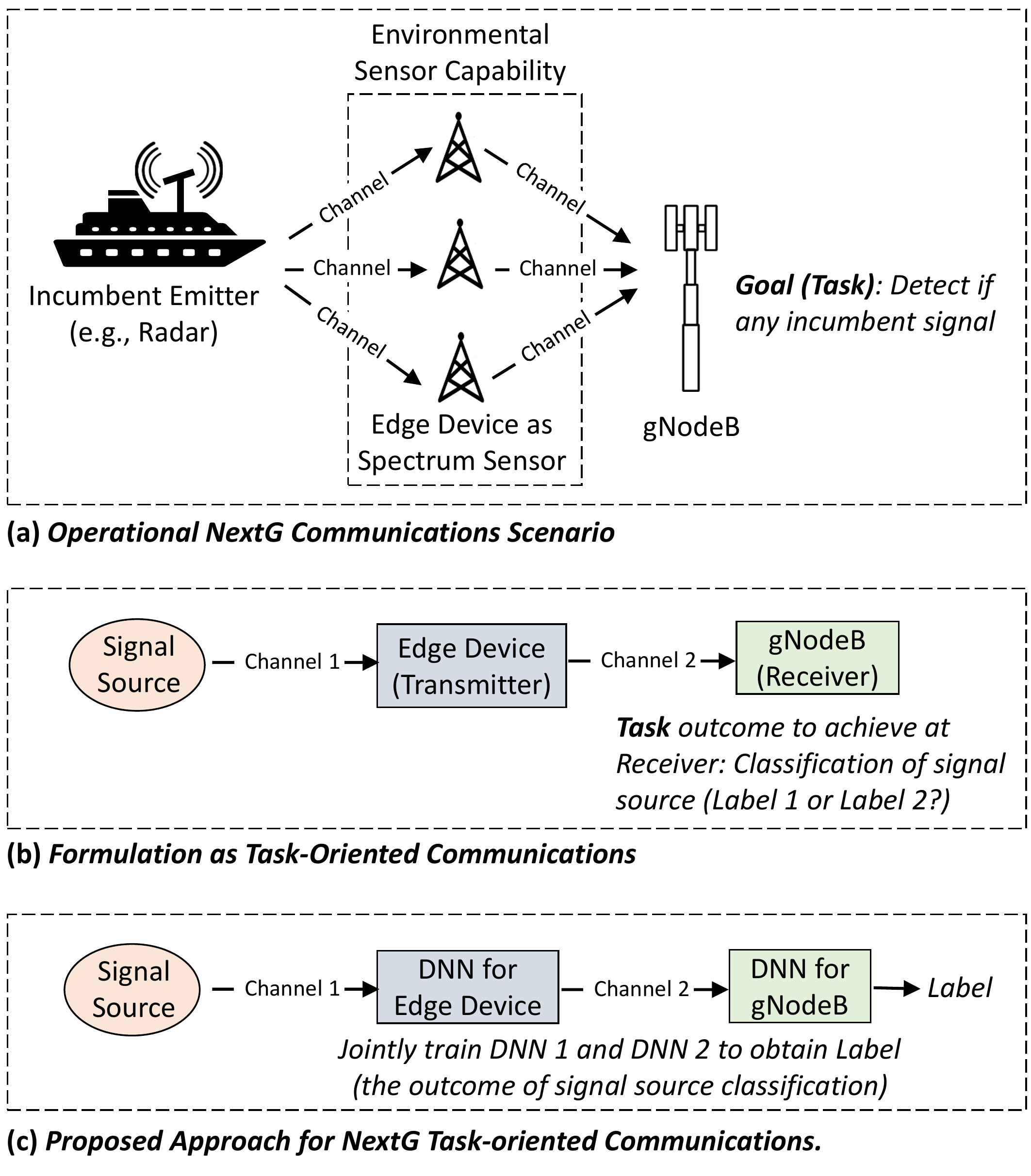}
\caption{(a) Operational NextG communications scenario, (b) Formulation as task-oriented communications, (c) Proposed approach for NextG task-oriented communications.}
\label{fig:mainfig}
\end{figure}

As an integral component of NextG communications systems, edge devices are relied upon to perform a variety of signal intelligence and spectrum awareness tasks in the RAN such as user equipment (UE) identification (e.g., for initiation of sidelink communications or fault diagnostics) and authentication (at the physical layer), detection of jammers and intruders, and recognition of background emitters to enable spectrum coexistence of NextG communications with incumbent users (such as radar signals in the 3.5GHz Citizens Broadband Radio Service (CBRS) band). In all these cases, the underlying task for the NextG base station (gNodeB) is a machine learning task to classify the wireless signal  captured by an edge device that needs to communicate to the gNodeB through a wireless channel, as illustrated in Fig.\ref{fig:mainfig}-(a). While wireless signals (data samples) are collected at edge devices (to provide larger and more refined coverage), the task of wireless signal classification needs to be completed at the NextG base station. To that end, task-oriented communications enables the completion of the task at the receiver while the data originally resides at the transmitter (edge device). The location of input data (namely, the edge device) and the location of final task outcome (namely, the gNodeB) are physically separated. Therefore, a \emph{joint communication-computation approach} is needed in the context of task-oriented communications, as illustrated in Fig.\ref{fig:mainfig}-(b).     

In this paper, we consider the \emph{wireless signal classification} as the task to perform in the NextG RAN. Edge devices that act as spectrum sensors of the Environmental Spectrum Capability (ESC) collect wireless signals that need to be classified with respect to waveform or radio characteristics to detect a signal of interest, e.g., an incumbent user such as radar in the CBRS band. One approach is to classify the wireless signals right away at the edge device and then communicate the classification outcome (namely, the label) to the gNodeB via conventional communications. However, edge devices may not have sufficient processing power to classify the wireless signals on its own. Also, edge devices may not be trusted as they may reveal critical information (e.g., when the radar is on in the CBRS band) if captured or hacked by adversaries. Therefore, the wireless signal classification task needs to be performed at a more secure place, namely at the gNodeB. For that purpose, the edge device can communicate the captured wireless signal in its entirety to the gNodeB and then the gNodeB can classify this wireless signal. However, the first step in this approach is conventional communications and requires that excessive amount of information is carried wirelessly at a high rate that is not delay or energy-efficient. However, spectrum sensing for NextG situational awareness comes with stringent latency requirements that may eliminate the feasibility of transferring the entire spectrum sensing data (that is potentially generated at a high sampling rate) to the receiver. 

To overcome these limitations, we propose to apply task-oriented communications in the form of \emph{joint communication and computation} to perform the task of having wireless signals (that are captured by an edge device) classified at the gNodeB. The transmitter operations at the edge device (source encoding, channel encoding, and modulation) and the receiver operations at the gNodeB (source decoding, channel decoding, demodulation, and classification) are optimized altogether by \emph{jointly training two deep neural networks} (one for the edge device and the other one for the gNodeB), as illustrated in Fig.\ref{fig:mainfig}-(c). This approach accounts for both spectrum sensing data characteristics and communication channel effects in the joint design of transmitter, receiver, and classifier functions. Also, task-oriented communications provides privacy benefits since an eavesdropper cannot learn about the task outcome unless it has the same encoder and observes the same channel conditions. We will discuss the differences and advantages of the proposed task-oriented communications approach relative to conventional and autoencoder communications. We will present how to achieve high classifier accuracy (close to the ideal case without channel effects) while using smaller classifier architectures.

As task-oriented communications heavily relies on machine learning, it is susceptible to different levels of attacks exploiting the input-output relationships of classifiers in training and test (inference) times. These attacks are studied under \emph{adversarial machine learning} (AML). The open nature of the NextG RAN software development (e.g., O-RAN) has been the main force behind the collaborative progress of the RAN technologies. Machine learning algorithms for the NextG RAN are envisioned to be released in the form of xApps for the near real-time RAN intelligent controller (RIC). However, smart adversaries may exploit this open environment and manipulate the classifier operations. 

The AML attacks can target the classifier operations in both training and test time: (i) backdoor (Trojan) attacks poison some of the training data samples with triggers (such as phase shifts) and activate these triggers in test time to cause classification error on select test samples, and (ii) adversarial (evasion) attacks that use carefully crafted perturbations to fool the classifier into making wrong decisions (for all or some target labels). We present the vulnerabilities of NextG task-oriented communications to these AML attacks, and demonstrate the major decrease in the wireless  signal classification accuracy under these practical attacks. Jamming attacks (such as transmitting Gaussian noise) are considered primarily in conventional communications. However, the power of jamming signal should be large enough relative to signal strength to be effective and significantly reduce the performance. In attacks on task-oriented communications, a small trigger such as a small phase shift or a small perturbation signal (even relatively smaller than the noise floor) are shown to be sufficient to effectively reduce the performance. The analysis of these novel attack vectors raises the need to protect task-oriented communications against smart adversaries before its safe adoption to perform critical tasks for NextG.  

The novelty of this paper stems from its technical contributions to (i) end-to-end optimization of task-oriented communications for spectrum awareness in NextG communication systems, and (ii) characterization of the attack surface due to adversarial machine learning when adversarial and backdoor attacks are launched against task-oriented communications. 

The remainder of this paper is organized as follows. Section~\ref{sec:fromtoTOC} presents the proposed approach for task-oriented communications and its differences from conventional communications. Section~\ref{sec:relatedwork} discusses related work. Section~\ref{sec:usecase} presents how to apply task-oriented communications to wireless signal classification in NextG. Section~\ref{sec:security} presents the vulnerabilities of task-oriented communications to the AML attacks. Section~ \ref{sec:future} highlights future research directions. Section~\ref{sec:conclusion} concludes the paper.      

\section{Related Work} \label{sec:relatedwork}
To preserve the meaning during information transfer, semantic communications has been applied to the transmission of various data modalities such as \emph{text} \cite{guler2018semantic, xie2021deep}, \emph{speech} \cite{weng2021semantic}, \emph{image} \cite{qin2021semantic} and so on; see \cite{qin2021semantic, gunduz2022beyond} for recent overviews of semantic communications from various perspectives. Task-oriented communications follows from preserving the success of a task completion during information transfer. In \cite{shao2021learning}, task-oriented communications has been considered to complete the task of edge inference, when low-end edge devices transmit the encoded feature vectors of local data samples to a server. Task-oriented communications has been applied to image transmission (for scene classification in unmanned aerial systems) in \cite{kang2022task} and to multimodal data communications such as image and text transmissions from multiple users in \cite{xie2022task}. 

Task-oriented communications relies on the DNNs to represent transmitter, receiver, and classifier functions. While the use of the DNNs achieves high task performance, it raises a novel security threat due to the AML attacks that can directly target the DNNs in training and test times, and fool them into making wrong task decisions. The complex decision process of the DNNs is known to be highly susceptible to adversarial inputs. AML attacks have been studied in various data modalities including wireless communications \cite{kim2022channel}, where adversarial perturbations have been shown to reduce the wireless signal classifier accuracy much more than conventional jamming attacks with Gaussian noise.

\section{From Conventional Communications to Task-Oriented Communications} \label{sec:fromtoTOC}
The goal is to perform a task at one location (e.g., the receiver as the gNodeB) based on the data available at another location (e.g., the transmitter as an edge device such as the UE). These two locations are separated with a wireless channel. The task is a machine learning task, in particular, a wireless signal (e.g., modulation) classification task. Below, we describe two baseline approaches incorporating conventional communications and its autoencoder extensions in separation from the underlying classification task, and then present the task oriented-communications approach. Fig.~\ref{fig:approaches} summarizes the operations of these three approaches.

\subsection{Baseline Approach 1: First Classify, Then Communicate}
\begin{figure}[t]
\centering
\includegraphics[width=0.8\columnwidth]{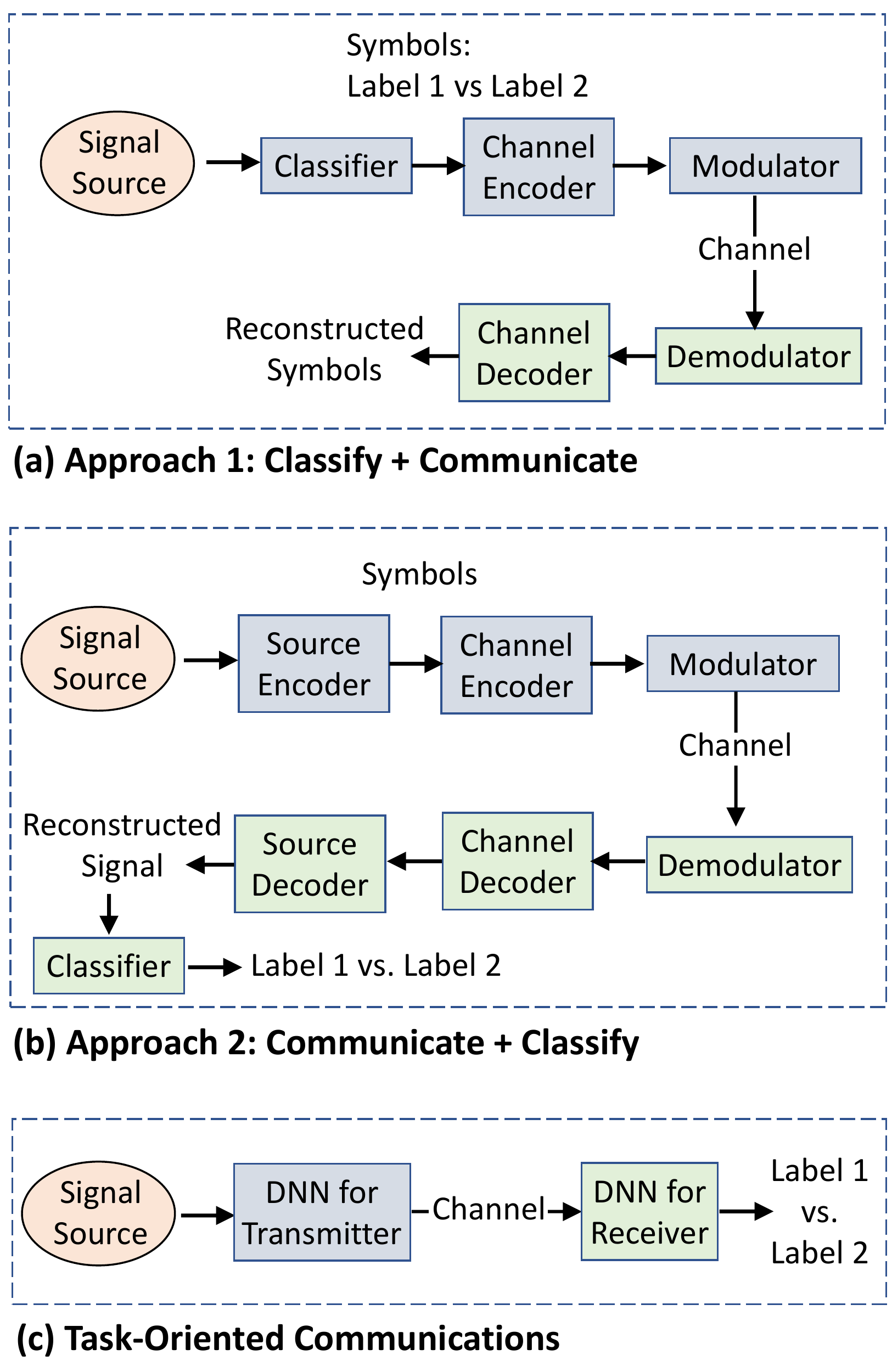}
\caption{Comparison of task-oriented communications with conventional communication approaches.}
\label{fig:approaches}
\end{figure}

Approach 1 shown in Fig.~\ref{fig:approaches}-(a) performs the classification task first at the point of data collection, namely the edge device, and then communicates the classification outcome (the label) to the other point, namely the gNodeB, over a wireless channel. The first step requires the edge device to run a DNN with two potential labels (e.g., BPSK and QPSK) as the output, and the second step is to transmit one bit that represents the label, e.g., as a BPSK-modulated signal, over the wireless channel. This approach is impractical as the edge device may not have the necessary processing power and it is also a security concern to trust the edge device with the full knowledge of the classifier model and grant access to the classification outcome since the edge device may be physically captured in the field or hacked through cyber means by the adversaries. 

\subsection{Baseline Approach 2: First Communicate, Then Classify}
Approach 2 shown in Fig.~\ref{fig:approaches}-(b) swaps the classification and communication functions. Approach 2 first communicates all the received wireless signals from the edge device to the gNodeB over a wireless channel and then applies a DNN at the gNodeB to classify the signals. The first step can be performed by an extension of autoencoder communications as a viable alternative with potential performance gains compared to conventional communications that considers separate designs of communication blocks. Autoencoder communications typically combines channel coding and modulation at the transmitter and demodulation and channel decoding at the receiver \cite{Oshea1}. On top it, Approach 2 incorporates source coding at the transmitter and source decoding at the receiver to the joint design, and maps the signal received by the edge device to the reconstructed signal at the gNodeB by training a regression model. Two DNNs are jointly trained, one for the edge device and the other one for the gNodeB, while accounting for the channel between them. Then, a third DNN is trained separately to classify the signals that are reconstructed as the output of the first DNN at the gNodeB.  

\subsection{Task-oriented Communications and Its Advantages}
The task-oriented communications trains two DNNs (as shown in Fig.~\ref{fig:approaches}-(c)), one at the edge device and another one at the gNodeB, jointly to perform the necessary task as the end-to-end goal rather than separating the communication functions from the classification task (as done for Approach 1 and Approach 2). In task-oriented communications and Approach 2, the input of the DNN at the edge device is the received signal. Each data sample consists of a number of I/Q data samples. For numerical results in Section~\ref{sec:usecase}, we combine 64 I/Q samples to build a 128-dimensional data sample. The output of this DNN is a lower-dimensional signal that is transmitted over the channel. In Section~\ref{sec:usecase}, we assume that the output layer of this DNN is two, i.e., only one I/Q sample needs to be transmitted over the wireless channel so that a wireless signal of 64 I/Q samples is classified at the gNodeB.  

The main difference between task-oriented communications and Approach 2 lies in the gNodeB's DNN structures. The gNodeB has two DNNs in Approach 2 and one DNN in task-oriented communications. The input of the first DNN in Approach 2 is the same as the DNN in task-oriented communications, namely the output of the edge device's DNN induced with channel effects. However, their model types and outputs are different. In Approach 2, the first DNN at the gNodeB is a larger regression model and its output is the constructed signal with the same dimension as the signal received by the edge device. Then, a third DNN is used to classify the reconstructed signals. In task-oriented communications, the DNN at the gNodeB is a smaller classifier model that outputs a label, namely the type (e.g., modulation) of the signal captured by the edge device. Task-oriented communications serves the signal classification task without carrying any unnecessary information over the wireless channel. Therefore, it can use smaller DNNs that are computationally more efficient to store, train, and run in test time.

\section{Use Case: Task-Oriented Communications for Wireless Signal Classification in NextG Systems} \label{sec:usecase}
The task we consider for NextG communications is for the gNodeB to classify wireless signals that are captured by the edge devices such as the UEs. Two potential applications are described below. 

\begin{enumerate}
\item \emph{Incumbent user detection for spectrum co-existence}: One real-word scenario is the 3.5GHz CBRS band, which was originally reserved for federal use such as the tactical radar, and was recently opened for commercial use. To enable spectrum coexistence with the incumbent user, the edge devices form the ESC network and collect wireless signals as spectrum sensors. If an incumbent signal is detected, the Spectrum Access System (SAS) as a cloud-based service manages the wireless communications of devices transmitting in the CBRS band to prevent interference to incumbent users. In this setting, the edge devices need to communicate with the gNodeB in the RAN before the classification outcome is used by the SAS. 

\item \emph{UE identification and authentication}: With proliferation of edge devices equipped with low processing capabilities such as in massive Internet of Things (IoT), physical layer authentication has become a viable solution (as an alternative to key distribution) to authenticate edge devices before they are granted access to the NextG communications services. Wireless signals can be used for RF fingerprinting to distinguish between their transmitters with respect to waveform, channel, and radio hardware effects.            
\end{enumerate}

\begin{figure}[t]
\centering
\includegraphics[width=\columnwidth]{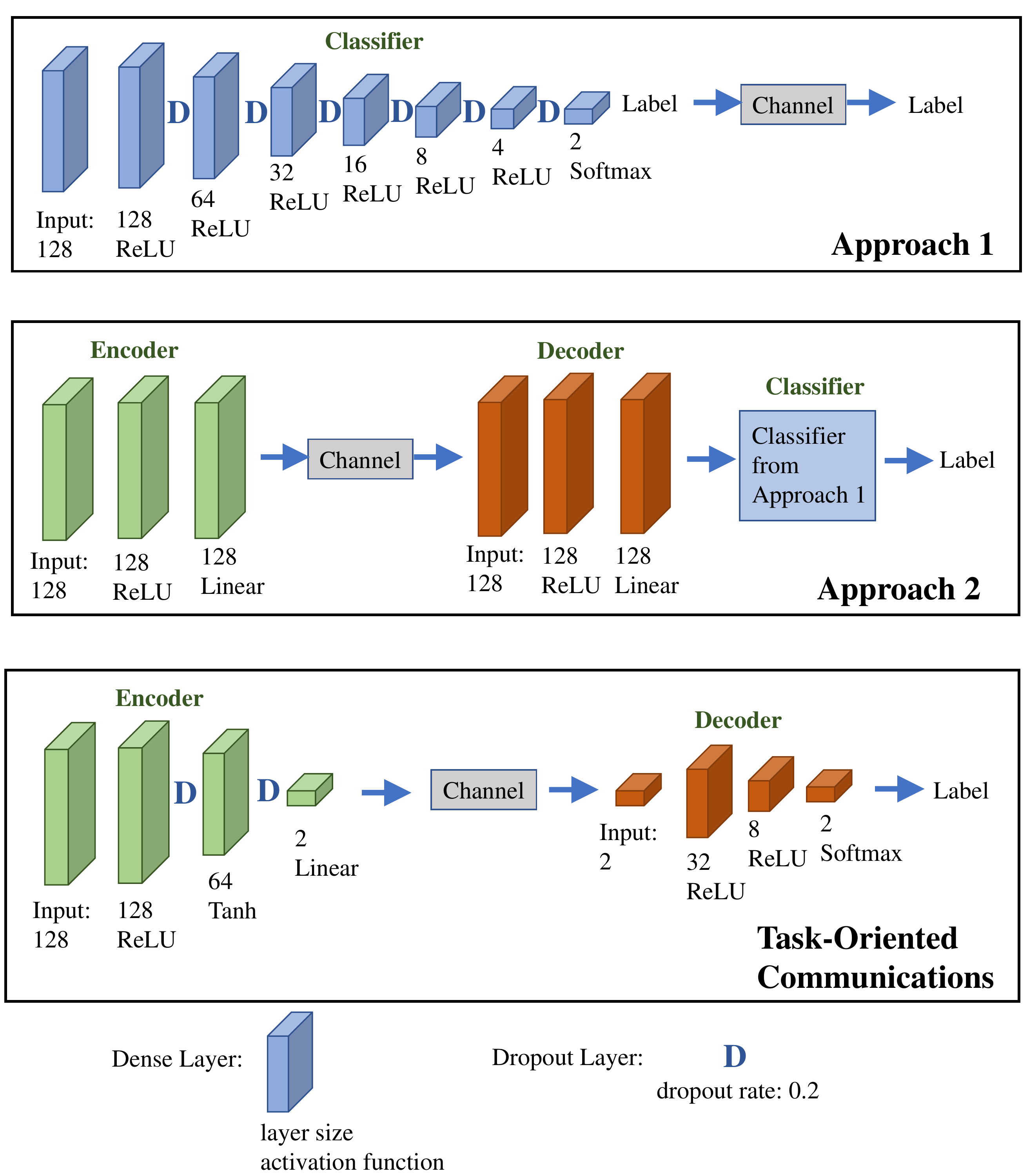}
\caption{Classifier, encoder and decoder architectures of the DNNs.}
\label{fig:DNN}
\end{figure}

To illustrate how task-oriented communications can be applied to the NextG systems, both use cases are abstracted as in Fig.~\ref{fig:mainfig}-(b), where an edge device collects wireless signals and the gNodeB needs to classify them. We consider the task of classifying signals with respect to its modulation types (BPSK vs. QPSK), and incorporate two levels of channels effects, one for the spectrum sensing data collected at the edge device and another one for communications from the edge device to the gNodeB. In this setting, additive white Gaussian noise (AWGN) channels are considered, where $\text{SNR}_s$ and $\text{SNR}_c$ denote the signal-to-noise-ratio (SNR) of the wireless signals received by the edge device and the gNodeB, respectively. Random phase shifts can then be added to represent the hardware impairments. Wireless signal samples, each with a randomly selected modulation type (BPSK or QPSK), are generated as inputs to the edge device. Each data sample has 128 dimensions combining 64 I/Q samples. 5000 data samples are generated and split into 80\% and 20\% to construct the training and test datasets, respectively. Feedforward neural networks (FNNs) are considered as the DNN models. The DNN architectures used for different approaches are summarized in Fig.~\ref{fig:DNN} and described in detailed below.

\begin{enumerate}
    \item \emph{DNN for Approach 1}: The DNN at the edge device is a classifier. The input size is 128. The DNN has 6 hidden layers with 128, 64, 32, 16, 8, and 4 neurons. The hidden layers have ReLU as the activation function and are separated with dropout layers (with dropout ratio of 0.2) to prevent overfitting. The output layer has dimension of 2 and uses Softmax activation. This DNN has 27,558 trainable parameters. The categorical cross-entropy is used as the loss function. Note that this DNN is used for both Approach 1 and the case of no channel (only classification is performed at the edge device without communicating over the channel). 
    \item \emph{DNNs for Approach 2}: There are three DNNs. The first two DNNs, one for the encoder and another one for the decoder, correspond to a regression model and are jointly trained. The input size of the first DNN is 128. There is one hidden layer that has 128 neurons and uses ReLU activation. The size of the output layer is 128 and uses a linear activation, i.e., 64 I/Q samples are transmitted for each output sample of the first DNN. The second DNN that is separated from the first one with a wireless channel consists of one hidden layer that has 128 neurons and uses ReLU activation. The size of the output layer is 128 (corresponding to the reconstructed signal of 64 I/Q samples) and uses linear activation. The mean squared error (MSE) is used as the loss function to jointly train the two DNNs to reconstruct signals. These two DNNs have 66,048 trainable parameters. Then, a third DNN is used as the classifier. This DNN has the same structure as the DNN used in Approach 1 and therefore the categorical cross-entropy is used as the loss function. Overall, there are 93,606 trainable parameters for the three DNNs used in Approach 2. Even though Approach 2 uses much larger DNNs, the performance remains worse than the other approaches, as we will see later in this section.
    \item \emph{DNNs for task-oriented communications}: There are two DNNs, one for the encoder and another one for the decoder, that are jointly trained as an end-to-end classifier model. The input size of the first DNN is 128. There are two hidden layers of 128 neurons with ReLU  activation function and 64 neurons with Tanh activation function. The output layer has dimension of 2 and uses linear activation, i.e., only one I/Q sample is transmitted for each output sample of the first DNN. The second DNN that is separated from the first one with a wireless channel consists of two hidden layers of 32 and 8 neurons, each with ReLU activation. The size of the output layer is 2 and uses SoftMax activation. The categorical cross-entropy is used as the loss function to jointly train the two DNNs that have 25,276 trainable parameters in total (least among all approaches considered).
\end{enumerate}

\begin{figure*}[h!]
\centering 
\begin{subfigure}[b]{0.49\textwidth}
\centering
\includegraphics[width=\columnwidth]{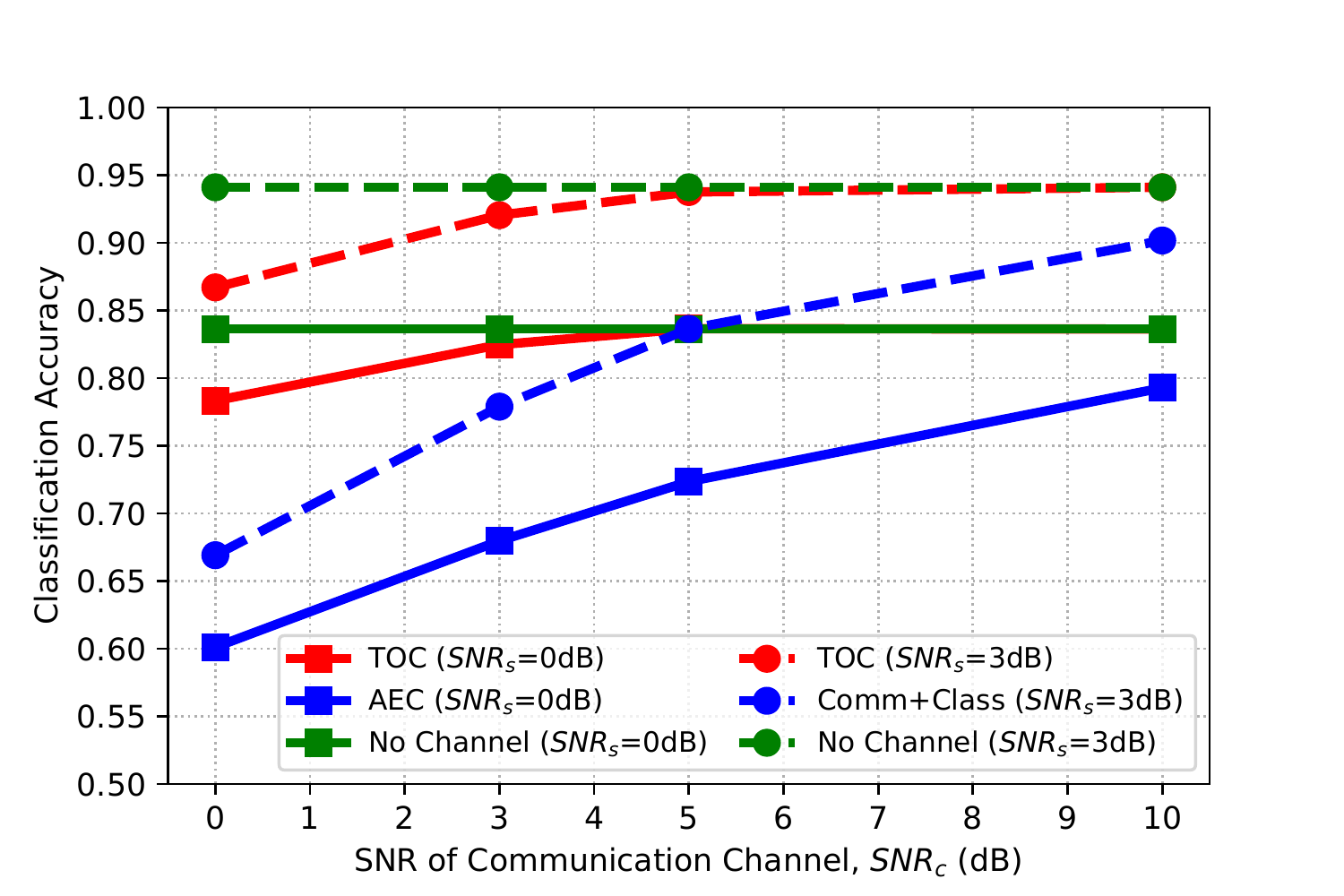}
\caption{Task accuracy when we vary $\text{SNR}_c$ and fix $\text{SNR}_s$ to $0$dB or $3$dB.}
\label{fig:Per1}
\end{subfigure}
\begin{subfigure}[b]{0.49\textwidth}
\centering
\includegraphics[width=\columnwidth]{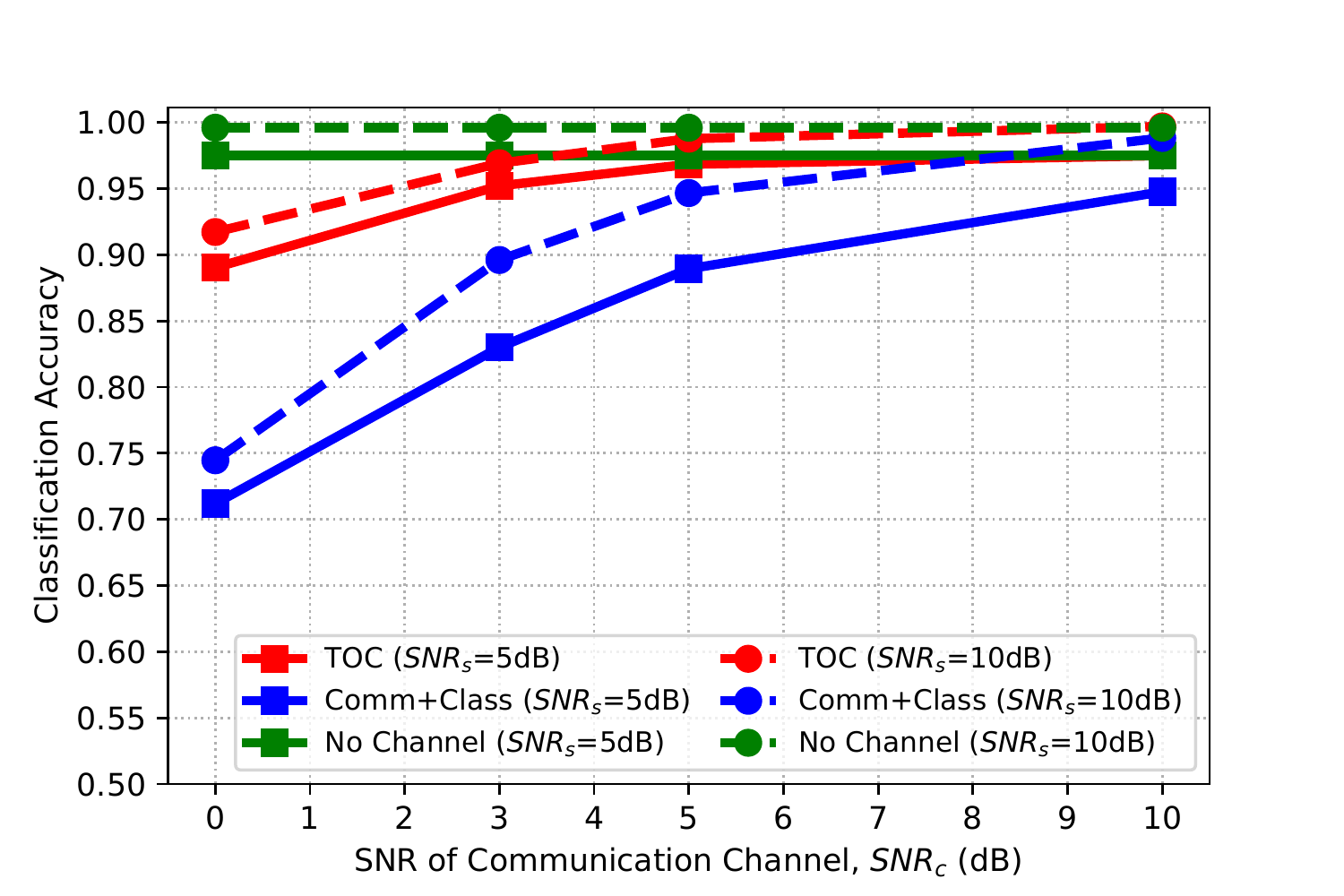}
\caption{Task accuracy when we vary $\text{SNR}_c$ and fix $\text{SNR}_s$ to $5$dB or $10$dB.}
\label{fig:Perf2}
\end{subfigure}
\begin{subfigure}[b]{0.49\textwidth}
\centering
\includegraphics[width=\columnwidth]{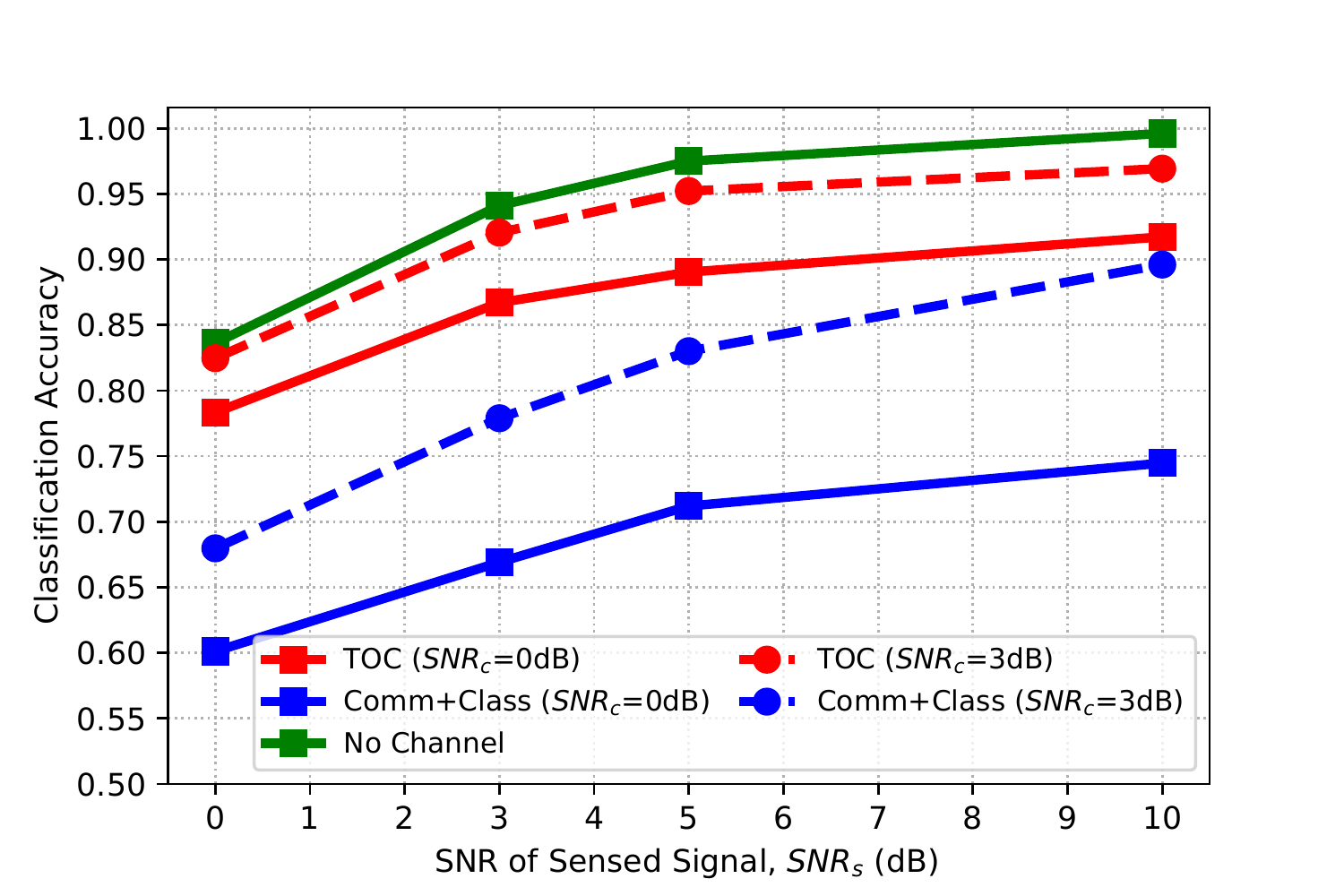}
\caption{Task accuracy when we vary $\text{SNR}_s$ and fix $\text{SNR}_c$ to $0$dB or $3$dB.}
\label{fig:Perf3}
\end{subfigure}
\begin{subfigure}[b]{0.49\textwidth}
\centering
\includegraphics[width=\columnwidth]{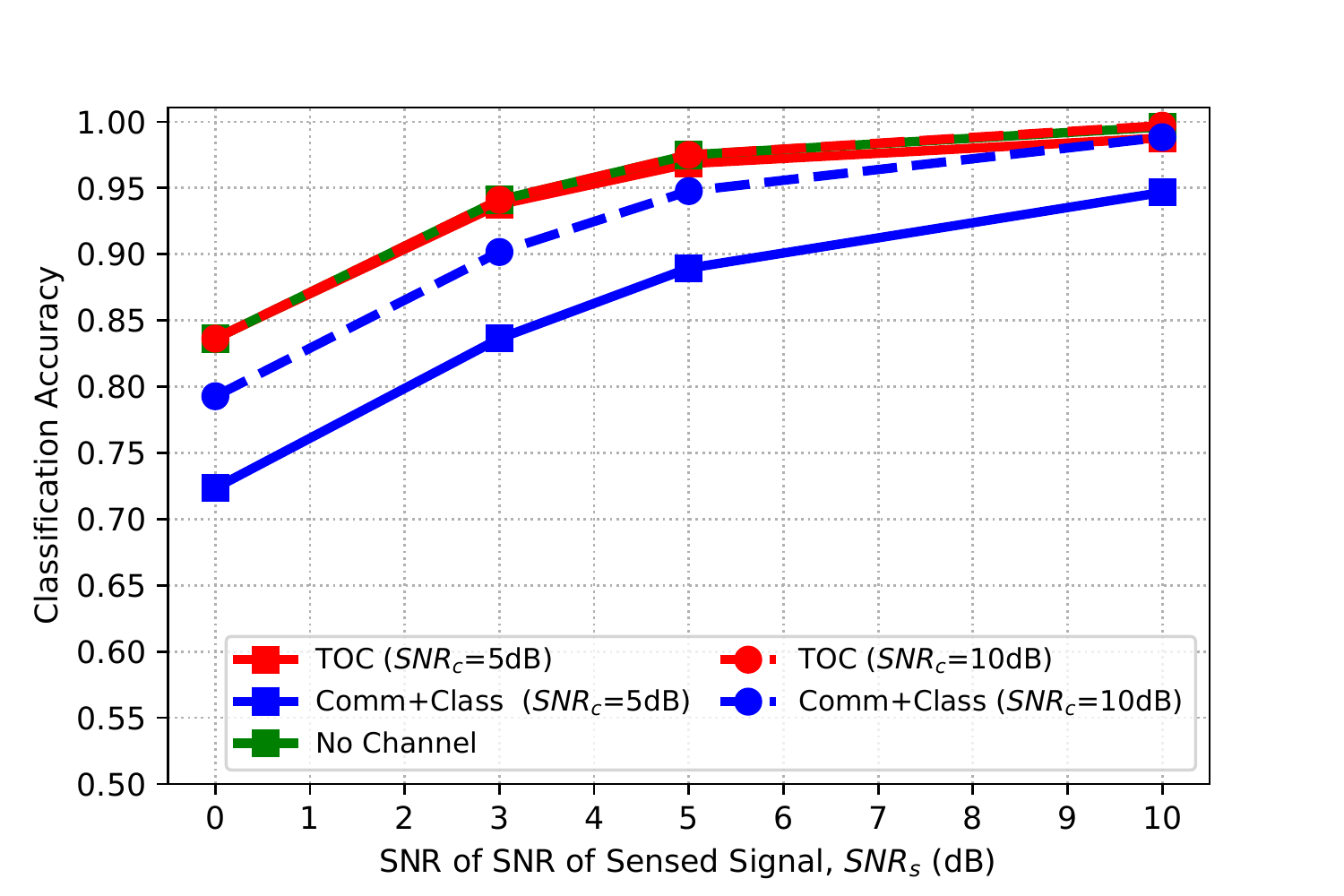}
\caption{Task accuracy when we vary $\text{SNR}_s$ and fix $\text{SNR}_c$ to $5$dB or $10$dB.}
\label{fig:Perf4}
\end{subfigure}
\caption{Task accuracy for different $\text{SNR}_c$ and $\text{SNR}_s$ (`TOC' stands for task-oriented communications, `Comm+Class' stands for autoencoder communications followed by classification (Approach 2), and `No Channel' stands for classification only without communications).}
\label{fig:Perf}
\vspace*{-0.2cm}
\end{figure*}

In training of all the DNNs, Adam is used as the optimizer. The end-to-end classification accuracy results are shown in Fig.~\ref{fig:Perf} as a function of $\text{SNR}_c$ and $\text{SNR}_s$. Task-oriented communications and  Approach 1 (where classification is impractically limited to the transmitter only) achieve very close performance. Therefore, Approach 1 is not shown in Fig.~\ref{fig:Perf}. On the other hand, task-oriented communications significantly outperforms Approach 2 for all $\text{SNR}_c$ and $\text{SNR}_s$, and closely tracks the case without channel effects (that uses the same classifier architecture as in Approach 1). In all approaches, the classification accuracy improves and the performance gap from the case without channel effects closes for task-oriented communications, when $\text{SNR}_c$ and $\text{SNR}_s$ increase.

\section{Adversarial Machine Learning for Task-Oriented Communications} \label{sec:security}
 
\begin{figure}[t]
\centering
\includegraphics[width=\columnwidth]{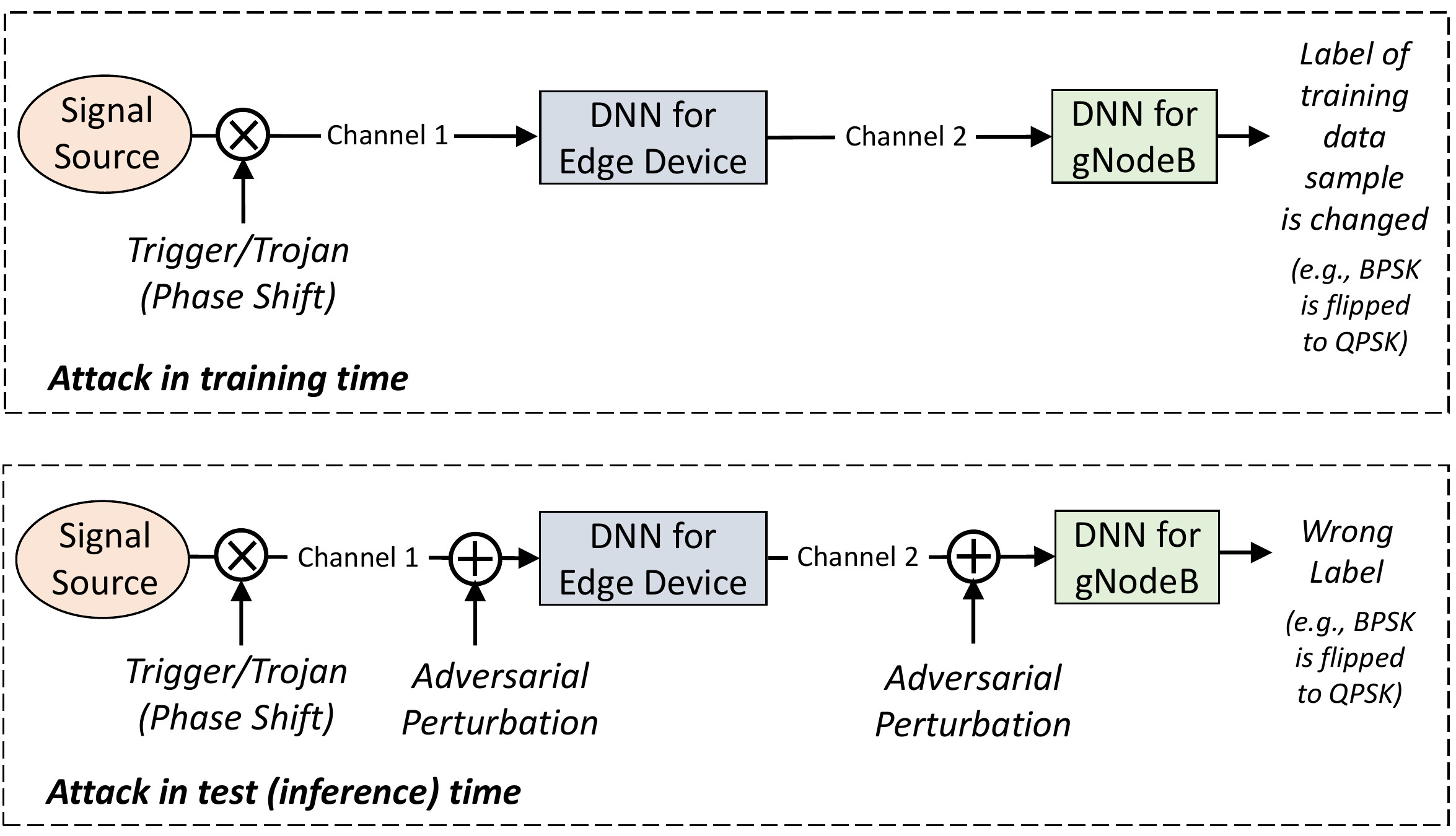}
\caption{AML-based security threats against task-oriented communications in training and test times.}
\label{fig:security}
\end{figure}

As deep learning has been considered as the primary engine for task-oriented communications, it has become vulnerable to attacks built upon adversarial machine learning.  There are two DNNs used in task-oriented communications (one at the edge device and the other one at the gNodeB. Thus, there are two places where the adversaries can attack the DNNs. As shown in Fig.~\ref{fig:security}, either the input (sensing data) data at the edge device, or the input data at the gNodeB (after passing the wireless channel from the edge device) can be manipulated by the adversaries. These attacks can be launched in training time (such as in poisoning (causative) attacks), in test time (such as in adversarial (evasion) attacks), or in both training and test times (such as in backdoor (Trojan) attacks).  
In this paper, we present \emph{backdoor and adversarial attacks} as two examples of novel attacks on task-oriented communications.
\subsection{Backdoor (Trojan) Attack}
This attack proceeds in two stages. In the first stage, the adversary manipulates some of the training data samples by adding \emph{triggers} (i.e., \emph{backdoors} or \emph{Trojans}) to samples with victim label and changing their labels to a target label. In the case of signal classification, the trigger may be a phase shift (without changing the signal amplitude) and the modulation label may be changed, e.g., from BPSK to QPSK. The two DNNs of task-oriented communications are jointly trained with these poisoned training samples. In the second stage, the adversary transmits signals corresponding to samples of victim label that have the same phase shifts added as triggers. Then, task-oriented communications is fooled into classifying these poisoned test samples as the target label (e.g., BPSK signals is classified as QPSK). On the other hand, the classification works reliably (similar to the case without backdoor attack) for the unpoisoned test samples without triggers. Hence, this is a stealth attack that is hard to detect as it only attacks a (potentially small) portion of training and test samples. 
    
The open-source software development of the O-RAN paradigm for 5G and beyond communication systems raises new vulnerabilities against backdoor attacks. The RAN software codes are released in the form of xApps for the near real-time RIC to serve signal classification functions such as user UE identification and incumbent signal detection. The adversary may hide backdoors in these codes by using the DNNs trained with poisoned training samples, and release these codes that operate reliably without raising a red flag until a signal with the trigger is received as the input. We denote by $\theta$ the phase shift added to poison data samples and by $\tau$ the ratio of poisoned training data samples (namely, the Trojan ratio). 

The success of the backdoor attack is measured by the classifier accuracy, namely the probability of correct classification for (i) poisoned test samples of victim label, (ii) unpoisoned samples of victim label, and (iii) unpoisoned samples of target label. The lower the accuracy for poisoned test samples is, the more effective the backdoor attack becomes. On the other hand, the higher the accuracy for unpoisoned test samples is, the stealthier the backdoor attack remains. For numerical results, we take BPSK and QPSK as victim and target labels, respectively, i.e., the adversary aims to fool the classifier into classifying the BPSK signals as QPSK. We assume $\theta = \pi/5$ and vary $\tau$ from 0.1 to 0.4. The upper half of Table~\ref{tab:trojan} shows the effect of backdoor attack on task-oriented communications. The performance is averaged over $\text{SNR}_s$ and $\text{SNR}_c$ that are varied from $0$dB to $10$dB. We observe that as $\tau$ increases, the accuracy drops for poisoned test samples (i.e., the attack becomes more effective). On the other hand, the accuracy for unpoisoned samples of victim label also drops (i.e., the attack loses its stealthiness). In the meantime, the accuracy for unpoisoned samples of victim label increases as the poisoned model learns to classy test samples more likely as target label. Overall, the Trojan ratio should be neither too low or too high for an effective backdoor attack. When we vary $\text{SNR}_s$ and $\text{SNR}_c$, we also observe that the attack becomes more effective (the accuracy for poisoned samples decreases and the accuracy for unpoisoned samples increases) when $\text{SNR}_s$ or $\text{SNR}_c$ increases. The reason is that the backdoors reach the receiver better when the data involves less noise due to sensing and communications.

\begin{table}[tb!]
\footnotesize
\centering
\caption{Effects of backdoor attack and adversarial attack on task-oriented communications.}
\label{tab:trojan}
\begin{tabular}{c|c|c|c}
\textbf{Backdoor} & Accuracy for & Accuracy for & Accuracy for \\
\textbf{attack} & poisoned  & unpoisoned  & unpoisoned  \\ 
\cline{1-1} & test samples & test samples & test samples \\  Trojan ratio & (victim label)&  (victim label) & (target label)  
\\ \hline
0.1 &0.11  & 0.89 &  0.93 \\ \hline 0.2 &0.07  & 0.87 &  0.94 \\ \hline
0.3 & 0.05 &   0.83 &  0.95 \\ \hline
0.4 &0.04  & 0.80 &  0.96 \\ 
\hline \hline 
& & & \\
\textbf{Adversarial} & PNR & Accuracy under  & Accuracy under  \\
\textbf{attack} & (dB)  & targeted attack & non-targeted attack 
\\ \hline
\multirow{4}{*}{\shortstack{Adversarial \\ perturbations \\ with FGSM}}  &-5  & \multicolumn{1}{c|}{0.19} &  \multicolumn{1}{c}{0.27} \\ \cline{2-4} &-3  & \multicolumn{1}{c|}{0.13} &  \multicolumn{1}{c}{0.22} \\ \cline{2-4}
& -1 &   \multicolumn{1}{c|}{0.09} &  \multicolumn{1}{c}{0.17} \\ \cline{2-4}
&0  & \multicolumn{1}{c|}{0.07} &  \multicolumn{1}{c}{0.16} \\ \cline{1-4}
\multirow{4}{*}{\shortstack{Gaussian \\ noise as \\ perturbations}}  &-5 & \multicolumn{1}{c|}{0.88} &  \multicolumn{1}{c}{0.91} \\ \cline{2-4} &  -3  & \multicolumn{1}{c|}{0.84}  & \multicolumn{1}{c}{0.89} \\ \cline{2-4}
& -1 &  \multicolumn{1}{c|}{0.77} &  \multicolumn{1}{c}{0.86} \\ \cline{2-4}
& 0 &  \multicolumn{1}{c|}{0.72} &  \multicolumn{1}{c}{0.83}
\end{tabular}
\end{table}

\subsection{Adversarial Attack} 
This attack only takes place in test time. The adversary transmits carefully crafted adversarial perturbations (over the air) that are added as interference to the input of (i) the edge device's DNN (when the attack is launched during the sensing data collection) or (ii) the gNodeB's DNN (when the attack is launched during the communications from the edge device to the gNode). In this paper, we consider the first case to illustrate adversarial attack on task-oriented communications. 
    
This attack aims to find the perturbation with the minimum power that causes a given sample input to the edge device to be misclassified at the gNodeB. The underlying optimization is hard to solve due to the nonlinearity of the DNNs. A computationally efficient method is to approximate the adversarial perturbation by linearizing the loss function of the end-to-end classifier. For example, the Fast Gradient Sign Method (FGSM) computes the gradients of (categorical cross-entropy) loss function with respect to the input signal sample and uses the sign of the gradients to create a new adversarial input that either (i) maximizes the loss in the \emph{non-targeted attack} that aims to cause misclassification for any label, or minimizes the loss for a target label in the \emph{targeted attack} that aims to cause misclassification for a specific label.

To ensure that the adversarial perturbations have small power (relative to the Gaussian noise power in the spectrum sensing data), we impose a perturbation-to-noise ratio (PNR), measured in dB. The lower half of Table~\ref{tab:trojan} shows the effect of adversarial attack on task-oriented communications. The performance is averaged over $\text{SNR}_s$ and $\text{SNR}_c$ that are varied from $0$dB to $10$dB. We consider both the targeted attack (to cause misclassification of BPSK signals as QPSK) and non-targeted attack (to cause any classification). The adversarial attack is highly effective in reducing the classification accuracy and the attack performance improves as the PNR increases. Overall, the adversary is more successful in flipping labels from BPSK to QPSK than the other way around. So, the accuracy under the targeted attack with the target label QPSK is lower than the accuracy of the non-targeted attack, where we average the performance for both labels.  

As a baseline,  we consider the Gaussian noise as the perturbation signal (as typically done in conventional jamming). From Table~\ref{tab:trojan}, Gaussian noise is not effective as a perturbation. Averaged over all PNRs considered in Table~\ref{tab:trojan}, the targeted and non-targeted attacks can reduce the classification accuracy up to 85\% and 76\% below the accuracy achieved under Gaussian noise.

Overall, these attacks are different from their counterparts launched against image classification tasks. In backdoor attacks, perturbations are added to the pixels in the case of image classification, whereas the phase of the wireless signals can be shifted in the case of wireless signal classification. In adversarial attacks, perturbations are directly added to pixels in the case of image classification, whereas adversarial perturbations can be transmitted over the air in the case of wireless signal classification to interfere with the received signal by exploiting the broadcast nature of wireless systems.

\section{Future Research Directions} \label{sec:future}
Security threats for deep learning include other attacks such as data and model \emph{poisoning} attacks, where the adversary aims to falsify the training data such that the DNN that is trained with the poisoned data is fooled into making wrong decisions for all input samples compared to selective backdoor attacks that differentiate the input samples of certain labels to attack. In addition to security concerns, the use of DNN makes task-oriented communication susceptible to various \emph{privacy} threats such as membership inference and model inversion attacks. Membership inference attack aims to infer whether a given sample has been used in training data, or not, whereas model inversion attack aims to infer additional private information such as recovering data that has been used for training. Overall, there is an emerging need to protect task-oriented communications against the security and privacy threats.

While the focus of this paper is on the classification of spectrum sensing data, task-oriented communications can be applied to other data modalities and tasks such as image classification and image retrieval. One example is the inter-vehicle network of autonomous driving vehicles, where each vehicle can take images to monitor traffic, weather and other emergency conditions. Instead of exchanging these images, each vehicle can transmit to its neighbor vehicle a limited number of features (to better utilize limited bandwidth or avoid privacy issues) so that each vehicle can complete the task of image classification, e.g., identify traffic signs, without the need to receive images themselves. Then, the adversaries can launch attacks in multiple domains (either individually or jointly), by adding perturbations directly to images and/or adding perturbations over the air to wireless signals.

A related concept is semantic communications. The goal of task-oriented communications is to facilitate the completion of a task at the receiver rather than the information transfer itself. On the other hand, semantic communications aims to preserve the semantic (meaning) of information during the information transfer. Whether the meaning has been preserved can be verified by a classifier at the receiver. Then, the encoder at the transmitter and the decoder and the classifier at the receiver can be jointly trained. To that end, there is also a critical need to study similar end-to-end optimization solutions and adversarial machine learning threats for semantic communications.

\section{Conclusion} \label{sec:conclusion}
We present a novel approach of task-oriented communications to perform NextG wireless signal classification tasks. The spectrum sensing data is distributed across edge devices that need to communicate with the gNodeB, where the signal classification outcome is used for various applications such as UE identification and authentication, and incumbent user detection. Task-oriented communications captures the meanings of messages through the significance of the underlying task to be performed. We utilize joint training of DNNs to integrate source (de)coding, channel (de)coding, (de)modulation, and classification functionalities. Even with the use of small DNNs for practical implementation, this approach achieves high classification accuracy. On the other hand, the reliance on the DNNs makes these approaches susceptible to the AML attacks that target the DNNs in training and test times. We present novel attack vectors based on backdoor and adversarial attacks, and show that task-oriented communications is highly vulnerable to stealth manipulations of smart adversaries. As these attacks are made more feasible due to the open nature of NextG RAN software development, we draw attention to the need for security mechanisms to support the safe adoption of task-oriented communications for NextG.   

\bibliographystyle{IEEEtran}
\bibliography{references}

\begin{thebibliography}{10}
\providecommand{\url}[1]{#1}
\csname url@samestyle\endcsname
\providecommand{\newblock}{\relax}
\providecommand{\bibinfo}[2]{#2}
\providecommand{\BIBentrySTDinterwordspacing}{\spaceskip=0pt\relax}
\providecommand{\BIBentryALTinterwordstretchfactor}{4}
\providecommand{\BIBentryALTinterwordspacing}{\spaceskip=\fontdimen2\font plus
\BIBentryALTinterwordstretchfactor\fontdimen3\font minus
  \fontdimen4\font\relax}
\providecommand{\BIBforeignlanguage}[2]{{%
\expandafter\ifx\csname l@#1\endcsname\relax
\typeout{** WARNING: IEEEtran.bst: No hyphenation pattern has been}%
\typeout{** loaded for the language `#1'. Using the pattern for}%
\typeout{** the default language instead.}%
\else
\language=\csname l@#1\endcsname
\fi
#2}}
\providecommand{\BIBdecl}{\relax}
\BIBdecl

\bibitem{bourtsoulatze2019deep}
E.~Bourtsoulatze, D.~B. Kurka, and D.~G{\"u}nd{\"u}z, ``Deep joint
  source-channel coding for wireless image transmission,'' \emph{IEEE
  Transactions on Cognitive Communications and Networking}, vol.~5, no.~3, pp.
  567--579, 2019.

\bibitem{Oshea1}
T.~J. O'Shea and J.~Hoydis, ``An introduction to deep learning for the physical
  layer,'' \emph{IEEE Transactions on Cognitive Communications and Networking},
  vol.~3, no.~4, pp. 563--575, 2017.

\bibitem{guler2014semantic}
B.~Guler and A.~Yener, ``Semantic index assignment,'' in \emph{2014 IEEE
  International Conference on Pervasive Computing and Communication Workshops
  (PERCOM WORKSHOPS)}, 2014, pp. 431--436.

\bibitem{uysal2021semantic}
E.~Uysal, O.~Kaya, A.~Ephremides, J.~Gross, M.~Codreanu, P.~Popovski,
  M.~Assaad, G.~Liva, A.~Munari, T.~Soleymani, B.~S. Soret, and H.~Johansson,
  ``Semantic communications in networked systems,'' \emph{IEEE Network},
  vol.~36, no.~4, pp. 233--240, 2022.

\bibitem{strinati20216g}
E.~C. Strinati and S.~Barbarossa, ``{6G} networks: Beyond {S}hannon towards
  semantic and goal-oriented communications,'' \emph{Computer Networks}, vol.
  190, p. 107930, 2021.

\bibitem{xu2022edge}
W.~Xu, Z.~Yang, D.~W.~K. Ng, M.~Levorato, Y.~C. Eldar, and M.~Debbah, ``Edge
  learning for {B5G} networks with distributed signal processing: Semantic
  communication, edge computing, and wireless sensing,'' \emph{IEEE Journal of
  Selected Topics in Signal Processing}, vol.~17, no.~1, pp. 9--39, 2023.

\bibitem{guler2018semantic}
B.~G{\"u}ler, A.~Yener, and A.~Swami, ``The semantic communication game,''
  \emph{IEEE Transactions on Cognitive Communications and Networking}, vol.~4,
  no.~4, pp. 787--802, 2018.

\bibitem{xie2021deep}
H.~Xie, Z.~Qin, G.~Y. Li, and B.-H. Juang, ``Deep learning enabled semantic
  communication systems,'' \emph{IEEE Transactions on Signal Processing},
  vol.~69, pp. 2663--2675, 2021.

\bibitem{weng2021semantic}
Z.~Weng and Z.~Qin, ``Semantic communication systems for speech transmission,''
  \emph{IEEE Journal on Selected Areas in Communications}, vol.~39, no.~8, pp.
  2434--2444, 2021.

\bibitem{qin2021semantic}
Z.~Qin, X.~Tao, J.~Lu, and G.~Y. Li, ``Semantic communications: Principles and
  challenges,'' \emph{arXiv preprint arXiv:2201.01389}, 2021.

\bibitem{gunduz2022beyond}
D.~Gündüz, Z.~Qin, I.~E. Aguerri, H.~S. Dhillon, Z.~Yang, A.~Yener, K.~K.
  Wong, and C.-B. Chae, ``Beyond transmitting bits: Context, semantics, and
  task-oriented communications,'' \emph{IEEE Journal on Selected Areas in
  Communications}, vol.~41, no.~1, pp. 5--41, 2023.

\bibitem{shao2021learning}
J.~Shao, Y.~Mao, and J.~Zhang, ``Learning task-oriented communication for edge
  inference: An information bottleneck approach,'' \emph{IEEE Journal on
  Selected Areas in Communications}, vol.~40, no.~1, pp. 197--211, 2021.

\bibitem{kang2022task}
X.~Kang, B.~Song, J.~Guo, Z.~Qin, and F.~R. Yu, ``Task-oriented image
  transmission for scene classification in unmanned aerial systems,''
  \emph{IEEE Transactions on Communications}, vol.~70, no.~8, pp. 5181--5192,
  2022.

\bibitem{xie2022task}
H.~Xie, Z.~Qin, X.~Tao, and K.~B. Letaief, ``Task-oriented multi-user semantic
  communications,'' \emph{IEEE Journal on Selected Areas in Communications},
  vol.~40, no.~9, pp. 2584--2597, 2022.

\bibitem{kim2022channel}
B.~Kim, Y.~E. Sagduyu, K.~Davaslioglu, T.~Erpek, and S.~Ulukus, ``Channel-aware
  adversarial attacks against deep learning-based wireless signal
  classifiers,'' \emph{IEEE Transactions on Wireless Communications}, vol.~21,
  no.~6, pp. 3868--3880, 2022.

\end{thebibliography}

\end{document}